\documentclass[a4paper]{jpconf}
\usepackage{graphicx}
\usepackage{textgreek}
\usepackage{amsmath}
\usepackage{amssymb}
\usepackage{placeins}
\usepackage{multirow}
\usepackage{circuitikzgit}
\usepackage{subfigure}
\usepackage{wrapfig}
\usepackage{capt-of}
\usepackage{booktabs}
\usepackage{varwidth}
\usepackage[noadjust]{cite}
\usepackage[breaklinks,colorlinks]{hyperref}

\definecolor{darkgreen}{rgb}{0.0, 0.2, 0.13}
\definecolor{darkblue}{rgb}{0,0,.5}

\hypersetup{pdftex=true, breaklinks=true, citecolor=darkblue, linkcolor=darkblue, menucolor=darkblue, pagecolor=darkblue, urlcolor=darkblue}

\begin{document}
\title{Development of signal readout electronics for LEGEND-1000}

\author{Frank Edzards$^{1, 2}$, Susanne Mertens$^{1, 2}$ and Michael Willers$^3$}

\address{$^1$ Max Planck Institute for Physics, Foehringer Ring 6, 80805 Munich, Germany}
\address{$^2$ Technical University of Munich, Arcisstrasse 21, 80333 Munich, Germany}
\address{$^3$ Lawrence Berkeley National Laboratory, 1 Cyclotron Rd, CA 94720, US}

\ead{edzards@mpp.mpg.de}

\begin{abstract}
The Large Enriched Germanium Experiment for Neutrinoless $\text{\textbeta\textbeta}$ Decay (LEGEND) is a future ton-scale experimental program that will search for neutrinoless double beta decay ($0\text{\textnu\textbeta\textbeta}$) in the isotope $^{76}$Ge with an unprecedented sensitivity. To achieve the projected discovery sensitivity at a half-life beyond $10^{28}\,$yr after about ten years of data taking, it is of utmost importance to operate in a background-free environment. This can be supported by using highly integrated readout electronics such as application-specific integrated circuit (ASIC) technology. In this work we investigated the properties and electronic performance of a commercially available ASIC charge sensitive preamplifier, the XGLab CUBE preamplifier, together with a p-type point contact high-purity germanium detector.
\end{abstract}

\section{Introduction}
The observation of neutrinoless double beta decay ($0\text{\textnu\textbeta\textbeta}$) would have major impacts on our understanding of the universe: It would shed light on the generation of neutrino mass and the smallness of the mass, open a window to understand matter dominance in our universe and provide information on the neutrino mass ordering. One of the most promising isotopes to search for the decay is germanium-76 ($^{76}$Ge). The newly formed LEGEND collaboration pursues a staged approach and plans to operate up to 1000\,kg of germanium detectors in the final stage of the experiment \cite{abgrall2017}. 
\\\\
The main readout electronics requirements for $0\text{\textnu\textbeta\textbeta}$ decay searches in the isotope $^{76}$Ge are a high energy resolution (low electronic noise), a small mass and high radiopurity, fast signal rise times, a good pulse shape analysis performance (discrimination of signal-like events from background events) and a large dynamic range. On the one hand, to obtain a high energy resolution and a good pulse shape discrimination performance, the readout electronics needs to be placed as close as possible to the detector. On the other hand, this is in contradiction with the radiopurity requirement, i.e.~any component in close proximity to the detector will contribute dominantly to the radioactive background. Consequently, a suitable trade-off between these requirements has to be found. A promising approach is to combine all relevant components in a single low-mass, low-background chip located very close to the detector using application-specific integrated circuit (ASIC) technology. 

\section{Signal readout electronics for LEGEND-1000}\label{ch:eletronics_legend}
One of the main challenges when scaling up a germanium-based $0\text{\textnu\textbeta\textbeta}$ decay experiment is the increased number of individual detectors and hence the increase of instrumentation, such as amplifiers, cables and connectors required. All these components are potential background sources and hence need to be of ultra-high purity and low mass. State-of-the art ASIC technology allows to integrate and miniaturize electronic components in a single chip, while being in no way inferior to discrete systems. The main advantages of using ASIC technology for $0\text{\textnu\textbeta\textbeta}$ decay searches compared to conventional amplifiers are a lower per-channel power consumption and a potentially higher per-channel radiopurity (e.g.~potentially fewer components, power cables, etc.). Furthermore, a better signal fidelity can be achieved since the ASIC technology allows for a high amplification gain close to the detector before sending the signal over a long distance to the data acquisition system. 

\section{Experimental setup}\label{ch:cube_facility}
\begin{wrapfigure}{R}{5cm}
\centering
\includegraphics[angle=0,width=0.25\textwidth]{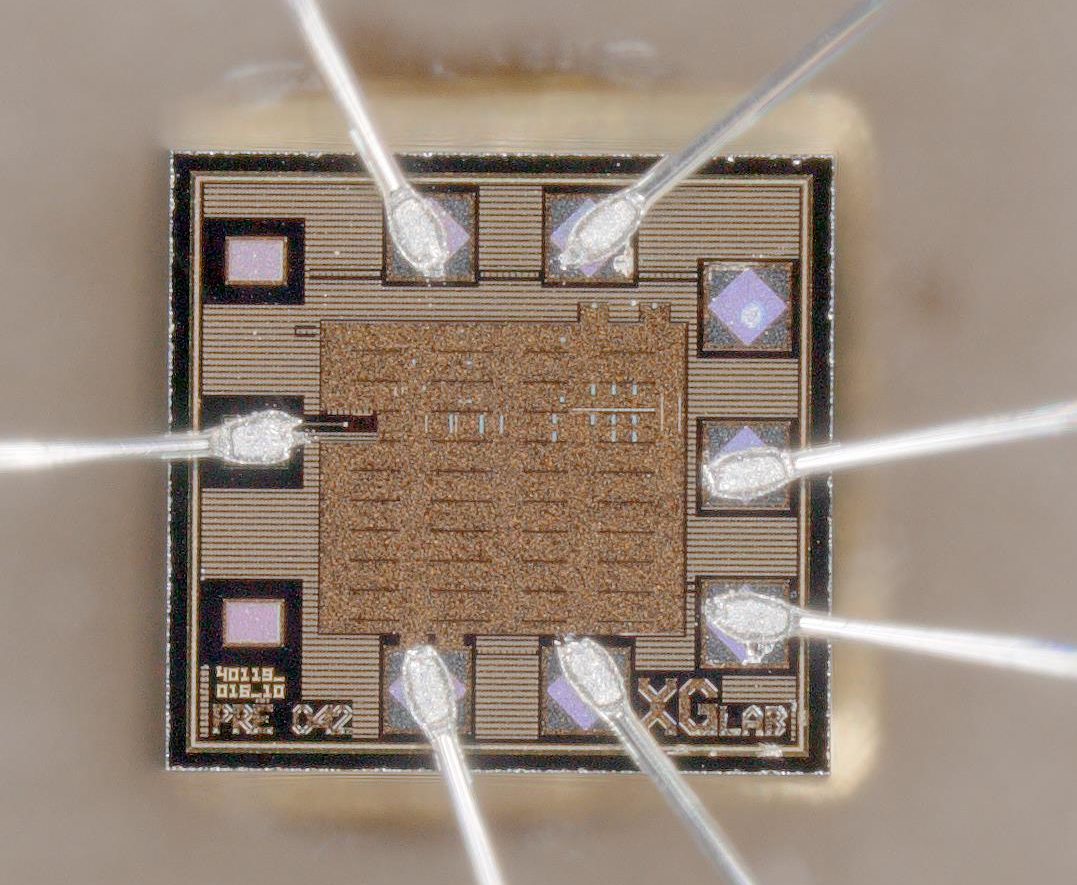}
\caption{Photograph of the CUBE ASIC.}
\label{graph:cube_asic}
\end{wrapfigure}
The CUBE$^3$ setup is a test facility that was designed to investigate signal readout electronics of high-purity germanium detectors for LEGEND. The core of the setup is a customized stainless steel vacuum cryostat that can host a germanium detector as well as the signal readout electronics. The pressure in the cooled system was stable at a level of $p\approx10^{-7}\,$mbar, even after several weeks without re-evacuating the cryostat. In addition, the temperature of the detector support structure was stable at a level of $T\approx98\,$K. For the investigation of an ASIC-based signal readout electronics, a p-type point contact (PPC) detector was used. The detector was instrumented with a commercially available CUBE ASIC obtained from the Italian company XGLab SRL \cite{bombelli2011}. The CUBE ASIC is a monolithic low-noise charge sensitive amplifier that has been developed in CMOS technology. Without being connected to a detector, the ASIC has a superior noise performance of 35.5\,e$^-$ (ENC) \cite{xglab2019}. \textbf{Fig.}~\ref{graph:cube_asic} shows a photograph of the chip. It is functional at cryogenic temperatures down to 50\,K and has a maximum power consumption of about 60\,mW. The internal feedback capacitance of $C_{\text{f}}=500\,\text{fF}\pm10\%$ translates into a large dynamic range corresponding to energies larger than 10\,MeV in germanium at cryogenic temperatures. The ASIC requires three supply voltages. Each of these supplies needs at least one bypass capacitor for the correct suppression of the voltage supply noise. 

\section{Results}\label{ch:results}
\begin{figure}[!h]
\vspace{-2.3cm}
\begin{center}
\mbox{
\hspace{-0.5cm}
\subfigure[$^{228}$Th energy spectrum.]{{\includegraphics[width = 0.40\textwidth]{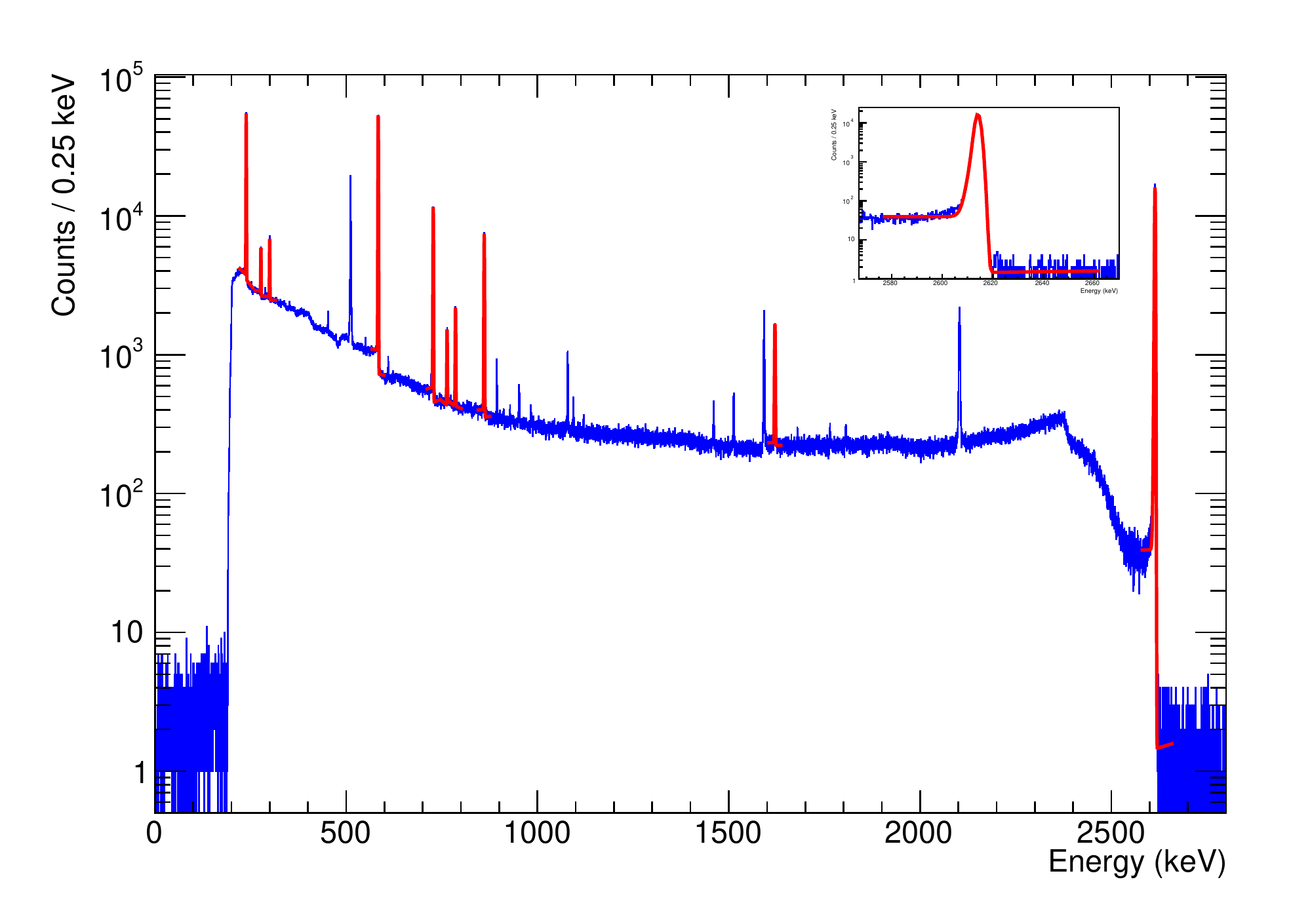}}}\quad
\subfigure[A/E scatter plot.]{{\includegraphics[width = 0.40\textwidth]{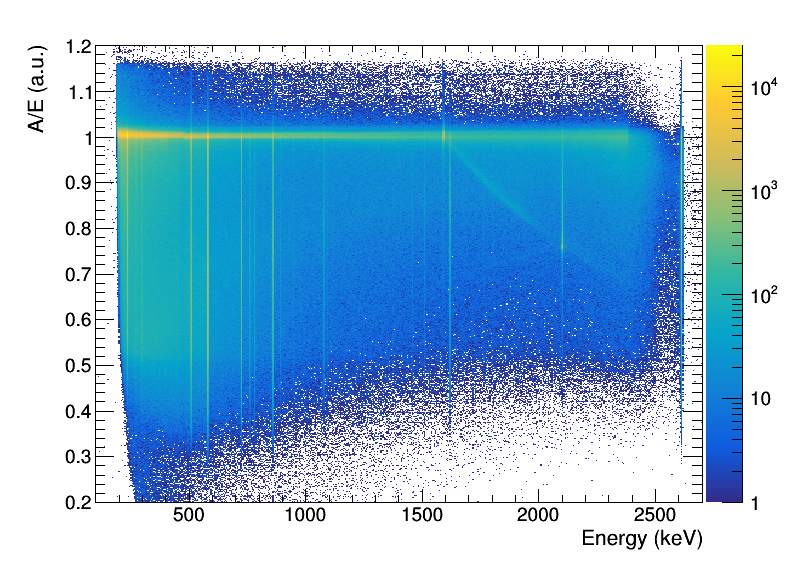}}}   
}
\caption{Energy spectrum (a) and normalized A/E distribution as a function of the energy (b) acquired during a $^{228}$Th calibration measurement.}
\label{graph:resolution_aoe}
\end{center}
\vspace{-0.5cm}
\end{figure}

First, the leakage current of the setup was investigated in dedicated measurements since it has a major impact on the electronic noise. The measurements showed a high stability at a level of about 16\,pA. Moreover, measurements without a detector at room temperature were carried out to measure the rise time of the CUBE preamplifier. Fast rise times as low as 15\,ns were obtained. The noise performance of the ASIC was investigated in terms of the baseline noise. A minimum baseline noise of about 537\,eV FWHM was obtained. To determine the energy resolution of the detector together with the ASIC-based readout system, it was irradiated with a strong collimated $^{228}$Th source. An example of the energy spectrum measured during a typical calibration run is shown in \textbf{Fig.}~\ref{graph:resolution_aoe}\,(a). An excellent energy resolution over a wide energy range was obtained. At the $Q_{\text{\textbeta\textbeta}}$-value (2039\,keV) and at the 2.6\,MeV $^{208}$Tl gamma peak, excellent energy resolutions of about 2.2\,keV FWHM and 2.5\,keV FWHM were obtained, respectively.
\FloatBarrier
\noindent
In order to fulfill the ultra-low background requirements for $0\text{\textnu\textbeta\textbeta}$ decay searches, it is indispensable to appropriately discriminate background events from signal-like events. One powerful background rejection method is based on the analysis of the shape of the signal pulses, commonly referred to as pulse shape discrimination (PSD). A commonly used discriminative quantity of the signal pulse shape is the ratio of the maximum amplitude of the current pulse $A$ and the amplitude (energy) of the charge pulse $E$: $A/E$ \cite{agostini2013}. The PSD performance of the CUBE ASIC together with the PPC detector (acceptance of background events in a $^{228}$Th calibration measurement: $\sim6\%$) was found to be comparable to the performance reported by the \textsc{Majorana Demonstrator} and \textsc{Gerda} experiments \cite{agostini2013, alvis2019}. An example for the normalized $A/E$ distribution is shown in \textbf{Fig.}~\ref{graph:resolution_aoe}\,(b). 

\vspace{-0.2cm}
\section{Conclusions}\label{ch:conclusion}
Signal readout electronics based on application-specific integrated circuit (ASIC) technology are ideally suited for low-background $0\text{\textnu\textbeta\textbeta}$ decay experiments like LEGEND. In this work we carried out a detailed investigation of the performance of a prototype ASIC. Therefore, a highly customized vacuum test facility, the CUBE$^3$ experiment, was designed. In the setup, the ASIC was operated together with a p-type point contact high-purity germanium detector. Dedicated measurements revealed that very fast signal rise times, low noise levels and an excellent energy resolution over a wide energy range can be obtained with an ASIC preamplifier. Moreover, the pulse shape discrimination capability of the ASIC-based readout system was demonstrated successfully. 

\section*{Acknowledgments}
This work was supported by the Max Planck Society, the Technical University of Munich and the DFG Collaborative Research Center "Neutrinos and Dark Matter in Astro- and Particle Physics" (SFB\,1258). The authors gratefully acknowledge support by the German Academic Scholarship Foundation (Studienstiftung des deutschen Volkes), the Alexander von Humboldt Foundation and the MPRG at TUM program.


\end{document}